\documentclass[fleqn,usenatbib]{mnras}

\usepackage[T1]{fontenc}
\usepackage{ulem}

\usepackage{graphicx}	
\usepackage{array}
\usepackage{mathtools}
\usepackage{amsmath}

\usepackage{newtxtext,newtxmath}

\title[TTV analysis of KELT-1\,b]{Transit timing variation analysis of the low-mass brown dwarf KELT-1\,b}

\author[Ba\c{s}t\"urk et al.]{%
\"O. Ba\c{s}t\"urk$^{1}$\thanks{E-mail: obasturk@ankara.edu.tr},
J. Southworth$^{2}$,
S. Yal\c{c}{\i}nkaya$^{1}$,
L. Mancini$^{3,4,5}$,
E. M. Esmer$^{1}$,
M. Tekin$^{6}$,
\newauthor
F. Tezcan$^{7}$,
D. F. Evans$^{2}$,
C. T. Tezcan$^{7,8}$,
I. Bruni$^{9}$,
and C. Ye\c{s}ilyaprak$^{7,8}$
\\\\
$^{1}$Ankara University, Faculty of Science, Astronomy \& Space Sciences Department, Tando\u{g}an, TR-06100, Ankara, T\"urkiye\\
$^{2}$\,Astrophysics Group, Keele University, Staffordshire, ST5 5BG, UK\\
$^{3}$\,Department of Physics, University of Rome ``Tor Vergata'', Via
della Ricerca Scientifica 1, I-00133, Rome, Italy \\
$^{4}$\,Max Planck Institute for Astronomy, K\"{o}nigstuhl 17,
D-69117, Heidelberg, Germany \\
$^{5}$\,INAF -- Osservatorio Astrofisico di Torino, via Osservatorio
20, I-10025, Pino Torinese, Italy \\
$^{6}$ Middle East Technical University, Faculty of Science, Physics Department, \"Universiteler mah, TR-06800, Ankara, T\"urkiye\\
$^{7}$ Atat\"urk University, Science Faculty, Department of Astronomy and Space Sciences, 25240, Erzurum, T\"urkiye\\
$^{8}$ Atat\"urk University Astrophysics Research and Application Center (ATASAM), Yakutiye, 25240, Erzurum, T\"urkiye\\
$^{9}$\,INAF -- OAS, Osservatorio di Astrofisica e Scienza dello Spazio di
Bologna, Via P. Gobetti 93/3, 40129, Bologna, Italy}

\date{Accepted XXX. Received YYY; in original form ZZZ}

\pubyear{2022}

\begin{document}
\label{firstpage}
\pagerange{\pageref{firstpage}--\pageref{lastpage}}
\maketitle

\begin{abstract}
  We investigate whether there is a variation in the orbital period of the short-period brown dwarf-mass KELT-1\,b, which is one of the best candidates to observe orbital decay. We obtain 19 high-precision transit light curves of the target using six different telescopes. We add all precise and complete transit light curves from open databases and the literature, as well as the available TESS observations from sectors 17 and 57, to form a transit timing variation (TTV) diagram spanning more than 10 years of observations. The analysis of the TTV diagram, however, is inconclusive in terms of a secular or periodic variation, hinting that the system might have synchronized. We update the transit ephemeris and determine an informative lower limit for the reduced tidal quality parameter of its host star of Q$_{\star}^{\prime} > (8.5 \pm 3.9) \times 10^{6}$ assuming that the stellar rotation is not yet synchronised. Using our new photometric observations, published light curves, the TESS data, archival radial velocities and broadband magnitudes, we also update the measured parameters of the system. Our results are in good agreement with those found in previous analyses.   
\end{abstract}




\section{Introduction}
\label{sec:introduction}
KELT-1\,b was the first low-mass companion found by the Kilodegree Extremely Little Telescope-North survey (KELT-North) for transiting planets \citep{siverd2012}. It is the closest known low-mass brown dwarf (BD) to its host star, orbiting every $P_{\rm orb} = 1.2175$\,d. At such a small orbital separation, tidal effects will lead to the transfer of angular momentum between the planetary orbit and the stellar spin. If the rate of the stellar rotation is longer than the orbital period, which is the usual case for giant planets in short-period orbits around cool stars, then the orbit is expected to decay and the star to spin up \citep{ogilvie2014}. KELT-1\,b, with its relatively large radius ($\sim1.14$ R$_{\rm Jup}$), high mass ($\sim27.7$ M$_{\rm Jup}$), and short period, should experience strong tidal interactions with its host star, making it a good candidate for the detection of orbital decay. However, its mid-F host star (effective temperature $T_{\rm eff} \sim6470$\,K) should have a thin convective envelope, which will not be efficient in dissipating the energy produced by the tidal interactions with the BD companion. This low efficiency will have an important effect on the properties of the system. Therefore, it is dynamically interesting to study the observable consequences of tidal interactions.

A lower energy dissipation efficiency is expected from the host star compared to that for cooler stars with thicker convective envelopes, which could prolong its survival in the very short-period orbit where it currently exists. This raises a question about the tidal quality parameter, which quantifies the dissipation of the tidal energy ($Q^{\prime}_{\star}$) by the host star. \citet{siverd2012} found that for $Q^{\prime}_{\star} = 10^8$, the orbital decay timescale should be less than 0.3\,Gyr, which is shorter than the age of the system. Based on a canonical value of $Q^{\prime}_{\star} = 10^6$ and the parameters of the system, the mid-transit time should shift by approximately 10 minutes after 10 years of observation with respect to a reference mid-transit time \citep{maciejewski2018}. \citet{baluev2015} analyzed the published transit light curves accumulated until the time of their study. They found no TTVs except for a tentative difference of 0.6\,s in the orbital period, and corrected the linear ephemeris as a result. \citet{maciejewski2018} also analyzed the published transit light curves in addition to their nine transit observations to reveal potential timing variations of the system. Although they did not find a statistically significant quadratic trend in the mid-transit times, they were able to constrain $Q^{\prime}_{\star}$ to have a lower limit of $8.4 \times 10^5$ at 95\% confidence based on the best fitting quadratic function, which is slightly lower than the canonical value they assumed  \citep{barker_ogilvie_2009}. \citet{maciejewski2022} recently strengthened this limit to $2.33^{+0.36}_{-0.38} \times 10^6$ based on new observations of KELT-1 from ground- and space-based facilities, including the Transiting Exoplanet Survey Satellite (TESS; \citealt{Ricker+15jatis}) mission data from sector 17.

The projected rotational velocity of the host star (corresponding to $1.329 \pm 0.060$\,d assuming spin-orbit alignment; \citealt{siverd2012}) points to synchronization with the orbital motion, the timescale of which is also expected to be short. This finding is in agreement with a suggested value of the rotation period, $P_{\rm rot} = 1.52 \pm 0.29$\,d, from \citet{vonEssen2021}. The authors found this as a residual frequency in the periodogram of the TESS light curve pre-whitened for the ellipsoidal variation, reflection, and Doppler beaming effects. However, the relatively hot host star is not observed to be very active. It therefore does not have large enough surface inhomogeneities to produce prominent flux variations that surpass the dominant reflection effect and the ellipsoidal variation, which would have helped us determine the rotation period unambiguously. Nevertheless, the system can be argued to be synchronized due to tidal interactions with its host star. If tidal equilibrium has been achieved, then the orbital period should be stable. However, the slight difference between the rotational periods derived from spectroscopy and spot-induced variations can be important given the  large uncertainties, indicating that the spin-up of the host star may not be complete and the system might not yet have reached tidal equilibrium.

KELT-1\,b is also intriguing in terms of its mass and formation mechanism, which further complicate its classification.  Because its mass of $27.7$\,M$_{\rm Jup}$ exceeds the mass limit for deuterium burning \citep{Spiegel++11apj}, it is classified as a low-mass BD. However, it is in the so-called BD desert \citep{grether2006} which is a scarcity of BD-mass objects at short-period orbits compared to planetary and stellar-mass objects. In fact, its mass is very close to the intersection of the opposite-slope mass functions for planetary and stellar companions at short-period orbits at $\sim30$\,$M_{\rm Jup}$, where the desert is driest \citep{grether2006}. The question is then whether it formed in the protoplanetary disk via the core accretion mechanism and migrated inwards, or formed directly by gravitational instability at a greater distance. Either case makes in-situ formation unlikely and increases the likelihood of a migration history. The M-dwarf companion to the KELT-1 system at 0$\arcsec.5$ (150 au if it is bound) might have played a role in this history too \citep{siverd2012}. \textit{Gaia} DR3 measurements indicate a faint object with a mean \textit{Gaia} magnitude of $G \sim21$ at 7\arcsec.26 separation with a large error bar due to its faintness. However, it is unlikely to be the object seen in the high angular resolution imaging because \textit{Gaia} is unable to detect objects with a large contrast between them and within 2\arcsec\ of each other \citep{brandeker2019,mugrauer2022}.

The circularization timescale is also short for KELT-1\,b's orbit. However, \textit{Spitzer} secondary eclipse observations \citep{beatty2014} are inconclusive in terms of orbital eccentricity. On the other hand, light curve solutions without the assumption of a circular orbit give a better fit for very small, but non-zero eccentricities \citep{vonEssen2021}. A potential perturber at a larger orbital period might be the source of such an eccentricity.

The mass and short orbital period of KELT-1\,b suggest that orbital decay should be observable on the timescale of a decade using data of typical quality from 1--2\,m class ground-based telescopes \citep{Birkby+14mn,maciejewski2018}. The presence of a potential perturber, a possible cause of the claimed non-zero eccentricity in the orbit of KELT-1\,b, can also be investigated in the same data. While the former effect will lead to a secular change in the orbital period, the latter will cause periodic changes in the arrival times of the transit signals of the planet due to the reflex motion about the common centre of mass with the perturbing body, via the Light-Time Effect (LiTE). These arguments prompted us to observe the system regularly in order to investigate potential variations in the transit timings due to both possibilities. We observed 19 transits of KELT-1\,b in total with different telescopes for ten years between 2012 and 2022. We also collected all the transit light curves from the Exoplanet Transit Database (ETD) and the literature, and downloaded the TESS 2-minute cadence light curves recorded during sectors 17 and 57 from the Mikulski Archive for Space Telescopes (MAST) web portal\footnote{https://mast.stsci.edu/portal/Mashup/Clients/Mast/Portal.html} of the Space Telescope Science Institute (STScI). We then analyzed all the light curves with {\sc exofast} \citep{eastman2013} and derived the mid-transit times to form a homogeneous dataset spanning more than ten years of observations for a thorough TTV analysis. 

The paper is organised as follows. We present our observations and data reduction scheme in Section~\ref{sec:observations}. The results of our analysis for the global modelling of the system, based on the most precise light curves, and for the transit timing are described in Section~\ref{sec:analysis}. The interpretation of the results is discussed in Section~\ref{sec:conclusion}.

\section{Observations and Data Reduction}
\label{sec:observations}

We observed nine transits of KELT-1\,b with the 1.23\,m telescope at the Calar Alto Observatory (CAHA) in Spain. We used the telescope-defocusing method to obtain a high photometric precision \citep{southworth2009}. The DLR-MKIII CCD camera gave a field of view (FoV) of $21\arcmin \times  21\arcmin.5$ at a plate scale of $0\arcsec.32$ per pixel. Six of the light curves were obtained through the $I_{\rm c}$ filter, two in $R_{\rm c}$, and one in $V$. 

Two transits were observed with the 1.52 m Cassini Telescope at the Astronomical Observatory of Bologna in Loiano (Italy) using a Gunn-$i$ filter. The plate scale of the CCD was $0\arcsec.58$ per pixel, the FoV was $13\arcmin \times 12\arcmin.6$, and the telescope was defocused. 

\begin{figure}
	\includegraphics[width=0.99\columnwidth]{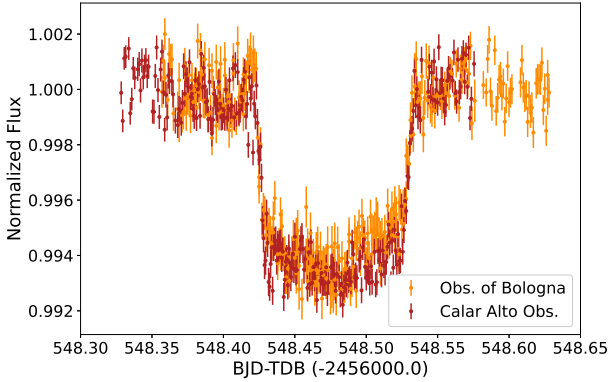}
    \caption{Normalized light curves recorded with the 1.52\,m telescope in Observatory of Bologna in Loiano (in orange) and 1.23\,m telescope in Calar Alto Observatory (in dark red) within Gunn$~ i$ and Cousins-I, respectively on 12 September 2013.}
    \label{fig:caha_loiano_20130912}
\end{figure}

One transit was observed with the 64-megapixel Bonn University Simultaneous Camera (BUSCA) on the 2.2\,m telescope at CAHA on 30 September 2016, simultaneously with the 1.23\,m telescope. Observations were obtained simultaneously through four filters: Str\"omgren $uby$ and Gunn-$z$. The data from the CAHA and Loiano observatories were reduced using the {\sc defot} code \citet{southworth2014}, which implements aperture photometry plus optimal weighting of comparison stars in the calculation of a differential-magnitude light curve. Unfortunately, the weather conditions were relatively poor so the ingress of the transit was not observed and the scatter in the remainder of the observations is higher than expected. The transit times measured from these data differ by almost 3\,min from the light curve from the 1.23\,m telescope so we did not include them in our TTV analysis. The light curve acquired during the same night using the 1.23\,m telescope is also not included, because it was noisy and we have much more precise light curves in the $R$-band. However, we did include the Gunn-$z$ data in the global modelling because it extends our wavelength coverage and the photometric precision is acceptable. The transit during the night of 12 September 2013 was also recorded simultaneously in these two observatories with Cousins-$I$ (CAHA) and Gunn-$i$ (Loiano) filters. We provide these light curves in Fig.~\ref{fig:caha_loiano_20130912} to illustrate the relative success of these observations in passbands with similar transmission functions.

We made use of the recently installed  \textit{Prof.\ Dr.\ Berahitdin Albayrak Telescope} (T80) at Ankara University Kreiken Observatory (AUKR) of Türkiye for four successful transit observations. The back-illuminated CCD with 1k pixels of 13 $\mu$m size provides a FoV of $11\arcmin.84 \times 11\arcmin.84$ when used with a focal reducer that reduces the plate scale to $53\arcsec.4$ per mm operated at the back of the 80 cm primary mirror. We employed SDSS-$g^{\prime}$ (two nights), SDSS-$r^{\prime}$ (one night), and SDSS-$i^{\prime}$ (one night) filters during the observations. 

We also observed a transit of KELT-1\,b with the 1 m Turkish telescope T100, located at T\"UB{\.I}TAK National Observatory of Türkiye (TUG) on the summit of Bak{\i}rl{\i}tepe mountain (altitude 2500\,m). The $f/10$ focal ratio provided a $21\arcsec~{\rm mm}^{-1}$ plate scale and an effective FoV of $21\arcmin \times 21\arcmin$. The readout from the $4096 \times 4096$ pixel CCD took 45\,s in the unbinned mode, which we employed due to the brightness of the star, allowing a good sampling of the 2.5\,hour-long transit, except for the ingress of the transit due to imperfect weather conditions just before the ingress. Therefore we had to exclude this light curve from the timing analysis. A Bessell $R$-band filter was used during the observation. 

Finally, two transits of the target were observed with a Johnson-$R$ filter with the 50\,cm Ritchey-Chr\'etien (RC) telescope (ATA50) of the Atat\"urk University's ATASAM Observatory in Erzurum, Türkiye, located at an altitude of 1824\,m. The Apogee Alta U230 CCD with $2048\times2048$ pixels, each of which has a size of 15\,$\mu$ m, provides a pixel scale of 0.77$^{\prime\prime}$ per pixel and an effective FoV of $13\arcmin \times 13\arcmin$. The data from the T80, T100 and ATA50 telescopes were reduced using the AstroImageJ (AIJ) \citep{collins2017} software package to obtain differential aperture photometry with respect to ensembles of comparison stars, yielding precise transit light curves for analysis in the current work. We used the observations from ATA50 and T80 only in the timing analysis, because we have more precise light curves in the same passbands which are more suitable for the global modelling.

KELT-1 was observed by TESS during sector 17 between 7 Oct 2019 and 2 Nov 2019, and sector 57 between 30 Sep 2022 and 29 Oct 2022. Both sectors were observed in short-cadence mode so have 2\,min integration times. These data were obtained from the MAST Portal. We examined the Target Pixel Files (TPF), Simple Aperture Photometry (SAP) and Pre-search Data Conditioning SAP (PDCSAP) light curves \citep{Jenkins+16spie}. We performed photometry with different apertures for both the target and the background, and compared our results with the SAP and PDCSAP light curves. Since \cite{vonEssen2021} studied the out-of-transit variability and the secondary eclipse in detail, we concentrated on the transit profiles to derive their mid-transit timings. We decided to use the light curves from the Data Validation Timeseries files (DVT), which are detrended from the correlated noise sources and hence stood out as the most convenient data type for our purpose. The visual companion at $\sim0\arcsec.5$, found by \cite{siverd2012} in high-resolution images, cannot be resolved in either TESS images or our own observations. It was found to be fainter by 5.6 mag in the $H$-band and 5.9 mag in the $K$-band so its flux contribution is less than 0.2\% at the optical wavelengths covered by our observations. This is well below the limit at which its contribution to the light curve is important \citep{southworth2020}.

We detrended some of our light curves that are affected by correlated noise. Instrumental effects, e.g.\ drifts in the x-y positions of stars on the CCD images due to imperfect telescope tracking and non-linearity issues, occasional clouds and other variability sources in weather conditions, all contribute to the red-noise budget for small (1\,m diameter or less) telescopes. Although we have not observed any obvious signatures of surface inhomogeneities induced by stellar activity on the raw light curves, which would affect their morphologies and eclipse depths, potential light curve asymmetries due to spot-crossing events in the transit chord may have also been smoothed out from the light curves we detrended. This is not very likely because KELT-1 is a mid-F type star with no indication of strong activity other than the tentative periodicity reported by \cite{vonEssen2021}. We employed a quasi-periodic kernel function for the stochastic part of a Gaussian Process (GP) accounting for both systematics and potential stellar variability. The deterministic part of the GP was defined by the base transit model. The centre and width of the Gaussian priors of the parameters in the transit models were set to the values derived by \citet{beatty2017}. The orbit was assumed to be circular. We normalized the out-of-transit flux to 1 for all the transit light curves. We employed a normal distribution for the white noise, which we controlled carefully and preserved its standard deviation to remove only the red noise component as efficiently as possible. In this manner, we detrended some of our light curves from correlated noise as described in \cite{yalcinkaya2021}, which we denoted with a ``d'' in the observing log (Table\,\ref{tab:observations}). We quantified the white and red (correlated) noise levels in all the light curves with the widely-used Photon Noise Rate (PNR) \citep{fulton2011} and $\beta$-factor \citep{winn2008} parameters, respectively and presented them in Table\,\ref{tab:observations}.

\begin{table*}
  \small
\centering
	\caption{Log of our transit observations of the transiting brown dwarf KELT-1\,b. Light curves denoted with the letter $d$ are detrended from the correlated noise and $^{*}$ were not used in the timing analysis since the ingress was not constrained well.}
	\label{tab:observations}
	\begin{tabular}{lcccccccccr} 
		\hline
		Telescope & Start date & Start time & End time & Filter & Exposure & Images & PNR & $\beta$ & Mid-Transit & Error \\
                 & UTC & UTC & UTC & & time (s) & Number & ppt &  & BJD$_{\rm TDB}$ & (d) \\ 
		\hline
                CAHA$^{\rm d}$ & 2012-09-09 & 00:14:22 & 04:54:54 & $I_{\rm c}$ & 145 & 113 & 0.239 & 1.386 & 2456179.576852 & 0.000177 \\
                CAHA & 2012-10-22 & 19:47:03 & 01:59:39 & $V$ & 120 & 163 & 0.489 & 2.956 & 2456223.407010 & 0.000447 \\
                CAHA & 2013-09-12 & 19:45:51 & 01:41:36 & $I_{\rm c}$ & 70 & 277 & 0.628 & 2.410 & 2456548.477686 & 0.000235 \\
                Loiano & 2013-09-12 & 20:26:58 & 02:57:11 & Gunn-$i$ & 85 & 309 & 0.678 & 2.500 & 2456548.477458 & 0.000255 \\
                CAHA & 2014-10-20 & 19:17:12 & 02:41:52 & $I_{\rm c}$ & 60 & 373 & 0.756 & 2.114 & 2456951.468633 & 0.000276 \\
                CAHA & 2015-08-20 & 00:04:13 & 04:33:41 & $I_{\rm c}$ & 60 & 225 & 0.751 & 1.556 & 2457254.624048 & 0.000303 \\
                CAHA & 2016-09-30 & 20:18:03 & 03:01:04 & $R_{\rm c}$ & 45 & 339 & 0.749 & 2.075 & 2457662.484192 & 0.000205 \\
                BUSCA$^{\rm{d},*}$ & 2016-09-30 & 20:42:05 & 03:00:31 & Gunn-$z$ & 150 & 122 & 0.330 & 3.005 & 2457662.484985 & 0.000526 \\
                Loiano$^{\rm d}$ & 2017-01-08 & 17:16:43 & 19:23:53 & Gunn-$i$ & 80 & 183 & 0.821 & 1.035 & 2457762.318983 & 0.000379 \\
                CAHA$^{\rm d}$ & 2017-08-04 & 22:03:15 & 02:11:01 & $I_{\rm c}$ & 90 & 120 & 0.475 & 1.606 & 2457970.511467 & 0.000252 \\
                CAHA & 2017-08-21 & 20:38:25 & 04:13:18 & $I_{\rm c}$ & 55 & 356 & 0.769 & 1.861 & 2457987.556687 & 0.000305 \\
                CAHA & 2017-09-23 & 19:18:25 & 02:00:36 & $R_{\rm c}$ & 60 & 335 & 0.530 & 1.641 & 2458020.427127 & 0.000200 \\
		T100$^{\rm{d},*}$ & 2019-10-29 & 16:06:22 & 21:22:33 & Bessell $R$ & 120 & 115 & 0.432 & 1.120 & 2458786.231872 & 0.000357 \\
                T80$^{\rm d}$ & 2021-07-29 & 20:27:35 & 00:20:08 & SDSS-$g^{\prime}$ & 150 & 91 & 0.557 & 1.171 & 2459425.416140 & 0.000541 \\
                T80$^{\rm d}$ & 2021-08-26 & 19:28:27 & 23:18:06 & SDSS-$g^{\prime}$ & 80 & 164 & 0.988 & 0.855 & 2459453.419241 & 0.000421 \\
                ATA50$^{\rm d}$ & 2021-09-28 & 16:12:41 & 21:30:37 & $R$ & 150 & 115 & 0.844 & 1.636 & 2459486.290677 & 0.000670 \\
                T80$^{\rm d}$ & 2021-10-10 & 20:59:35 & 01:59:22 & SDSS-$r^{\prime}$ & 80 & 197 & 0.851 & 1.236 & 2459498.465202 & 0.000370 \\
                ATA50$^{\rm d}$ & 2021-10-26 & 15:08:19 & 22:28:08 & $R$ & 150 & 142 & 0.505 & 1.664 & 2459514.293822 & 0.000503 \\
                T80$^{\rm d}$ & 2022-11-11 & 18:51:31 & 22:34:31 & SDSS-$i^{\prime}$ & 110 & 115 & 0.461 & 0.956 & 2459895.368370	& 0.000273 \\
		\hline
        \end{tabular}
\end{table*}

\section{Analysis and Results}
\label{sec:analysis}
\subsection{Global Modelling}
\label{subsec:global_modelling}
We selected the ``best'' transit light curves of KELT-1\,b, based on PNR and $\beta$-factor statistics, in different passbands to extend the wavelength coverage as much as possible for a global modelling to derive system and planetary parameters. Since we detrended some of our light curves, they have been corrected for spot-induced modulations, observational effects, and correlated noise due to instrumental effects and stellar variability. Because they have longer timescales than the transit duration, amplitudes far too small to be recovered by ground-based observations, and because we did not have a low-frequency component in the GP, the reflection effect, Doppler boosting, and ellipsoidal variations are already included in the white-noise budget. Although the detrended light curve in the SDSS-$r^{\prime}$ band acquired with T80 in Ankara is of high quality, there is a superior light curve in the same passband obtained in the University of Louisville Moore Observatory (ULMO) published in the discovery paper \citep{siverd2012}, which we used in the global modelling instead. We made use of two light curves (from BUSCA in Gunn-$z$ and T100 in Bessell $R$ during the global modelling although we discarded them from the timing analysis because the ingress was not constrained well due to poorer weather conditions at the time. While, light curves are analyzed separately to measure their mid-transit times, heavily influenced by the contact times of a transit, they are modelled as an ensemble to derive fundamental parameters of the system.

\begin{figure}
	\includegraphics[width=0.99\columnwidth]{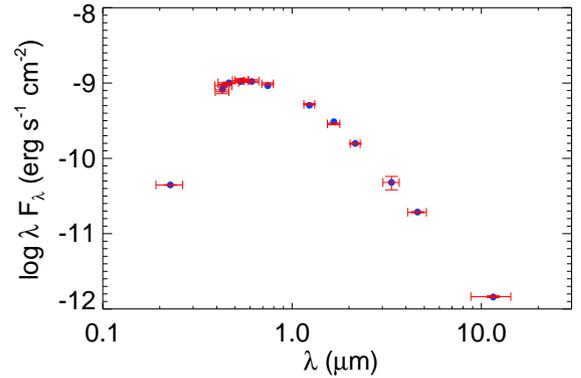}
    \caption{Broadband fluxes of KELT-1 (red data points with error bars) and model fluxes (blue points).}
    \label{fig:sed}
\end{figure}

We then collected broadband apparent magnitudes of the host star in different passbands used mostly in space-borne observations (see Table~\ref{tab:sed}) to fit the spectral energy distribution (SED) of the host star (Fig.~\ref{fig:sed}) in {\sc exofastv2} by using the Modules for Experiments in Stellar Astrophysics (MESA) Isochrones and Stellar Tracks (MIST) bolometric correction grid \citep{choi2016}. The stellar $(T_{\rm eff}$, metallicity ([Fe/H]) and surface gravity ($\log{g}$) were adopted from the most detailed spectroscopic analysis of the host star \citep{siverd2012} while the distance was calculated from the \textit{Gaia} EDR3 parallax \citep{gaia2016,gaia2021a,gaia2021b} with the offset given by \citet{Lindegren2021} and provided as a Gaussian prior during the SED fitting. The maximum value of the interstellar extinction ($A{_{V}}$) along the line of sight was limited to the value given by \citet{schlegel1998}. As a result, we obtained $T_{\rm eff}$ = $6491^{+45}_{-45}$~K, $T_{\rm eff,SED}$ = $6400^{+64}_{-56}$~K, $R_{\star}$ = $1.499^{+0.034}_{-0.032}\,$R$_{\sun}$ and $R_{\rm \star,SED}$ = $1.542^{+0.011}_{-0.012}\,$R$_{\sun}$ from the SED fitting. The values with the SED subscript comes purely from the SED analysis, i.e.\ without a Gaussian prior on $T_{\rm eff}$. We provided the empirically derived stellar radius from the best-fitting SED model (with the $T_{\rm eff}$ prior implied) as a Gaussian prior during global modelling to increase the accuracy of the absolute parameters of the system. The value of the Gaussian prior width for $R_{\star}$ was set to 3.5\% of its value as suggested by \citet{tayar2020}. We also provided the $T_{\rm eff}$ value given by \citet{siverd2012} instead of our SED value because the $T_{\rm eff}$ measurement from high-resolution spectra should be superior to the SED analysis.

\begin{table}
\centering
\caption{Broadband apparent magnitudes for KELT-1.}
\label{tab:sed}
\begin{tabular}{ccc}  
\hline
\hline
Passband & $\lambda_{\rm eff}\,(nm)$ & Magnitude \\
\hline
\multicolumn{3}{l}{APASS-DR9 \citep{henden2016}}\\
\hline
Johnson $B$ & 437.81 & $11.189 \pm 0.045$\\
Johnson $V$ & - & $10.659 \pm 0.045$\\
SDSS $g'$ & 464.04 & $10.916 \pm 0.022$ \\
SDSS $r'$ & 612.23 & $10.559 \pm 0.044$ \\
SDSS $i'$ & 743.95 & $10.441 \pm 0.038$ \\
\hline
\multicolumn{3}{l}{GALEX \citep{bianchi2017}}\\
\hline
galNUV & 227.44 & $15.089 \pm 0.013$ \\
\hline
\multicolumn{3}{l}{2MASS \citep{cutri2003}}\\
\hline
$J_{\rm 2MASS}$ & 1235 & $9.682 \pm 0.022$ \\
$H_{\rm 2MASS}$ & 1662 & $9.534 \pm 0.030$ \\
$K_{\rm 2MASS}$ & 2159 & $9.437 \pm 0.019$ \\
\hline
\multicolumn{3}{l}{All WISE \citep{cutri2013}}\\
\hline
WISE1 & 3352.6 & $9.414 \pm 0.220$ \\
WISE2 & 4602.8 & $9.419 \pm 0.020$ \\
WISE3 & 11560.8 & $9.386 \pm 0.034$ \\
\hline
\multicolumn{3}{l}{Tycho-2 \citep{hog2000}}\\
\hline
$B_T$ & 428.0 & $11.363 \pm 0.065$ \\
$V_T$ & 534.0 & $10.701 \pm 0.057$ \\
\hline
\end{tabular}
\end{table}

\begin{figure*}
	\includegraphics[width=0.85\paperwidth]{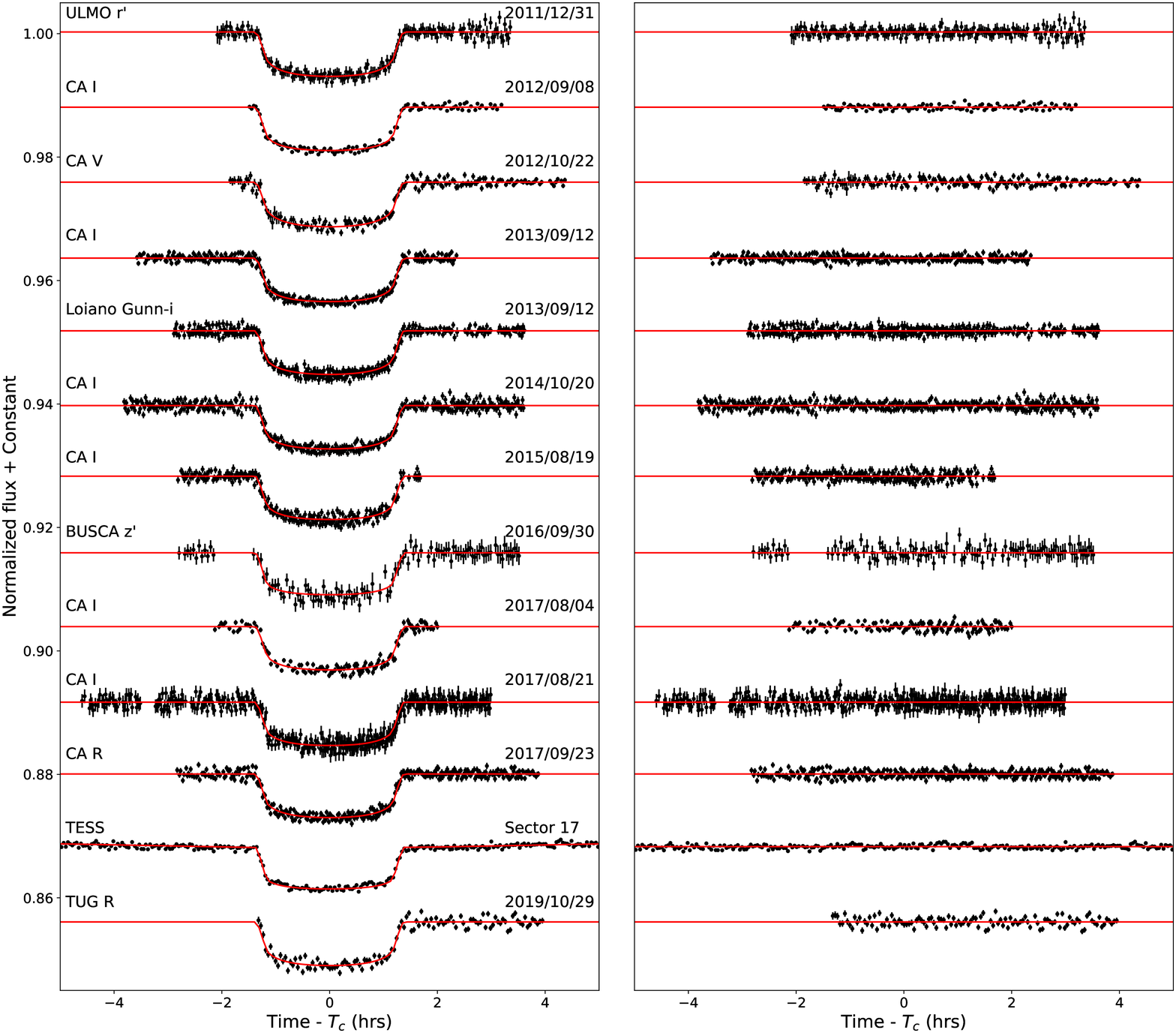}
    \caption{Our individual transit light curves and one light curve from \citet{siverd2012} (on top) and their {\sc exofastv2} models. Data points and their errorbars are in black, while their models are given with continuous red lines. CA: CAHA observations, TUG: T\"UB{\.I}TAK National Observatory of Türkiye, BUSCA: Bonn University Simultaneous CAmera observations, ULMO: University of Louisville Moore Observatory. The residuals from the models are given in the right-hand panel.}
    \label{fig:lcmodel}
\end{figure*}

\begin{figure*}
	\includegraphics[width=0.85\paperwidth]{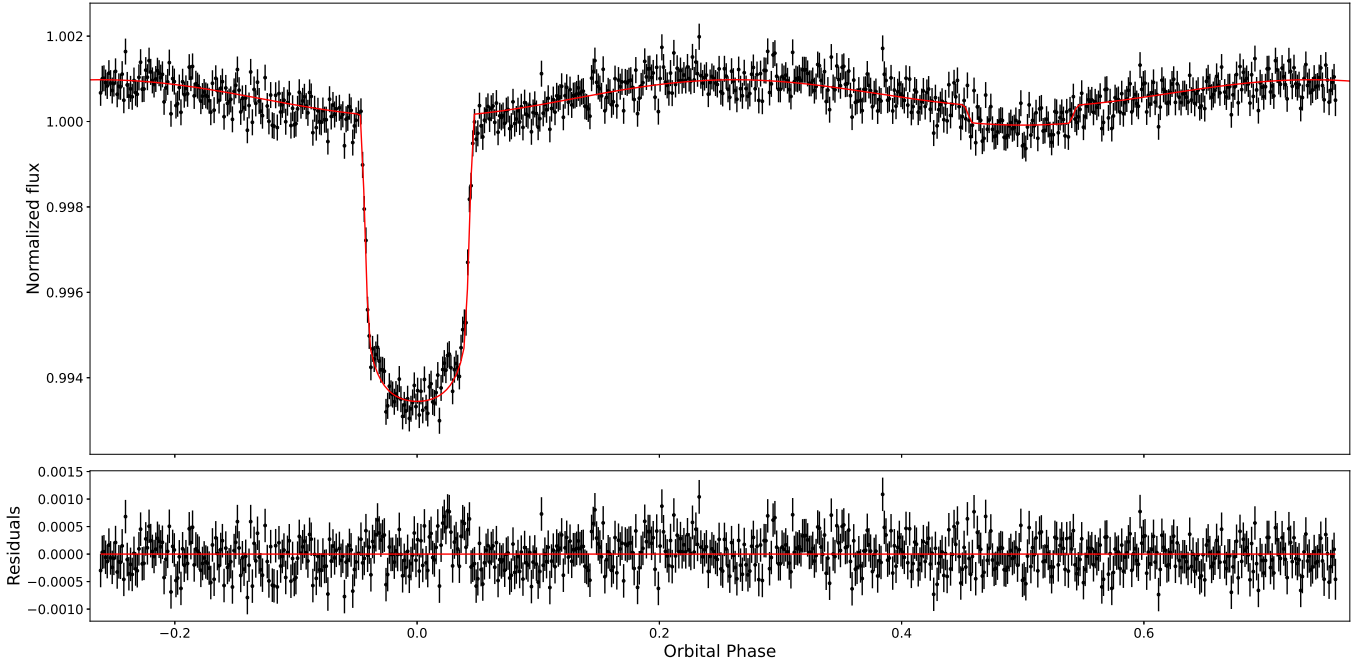}
    \caption{Phase folded TESS phase curve binned to 2 minutes from sector 17
    (black points with error bars) and the best-fitting global model (red continuous line). Residuals are shown in the bottom panel.}
    \label{fig:phasecurve}
\end{figure*}

\begin{figure}
	\includegraphics[width=0.995\columnwidth]{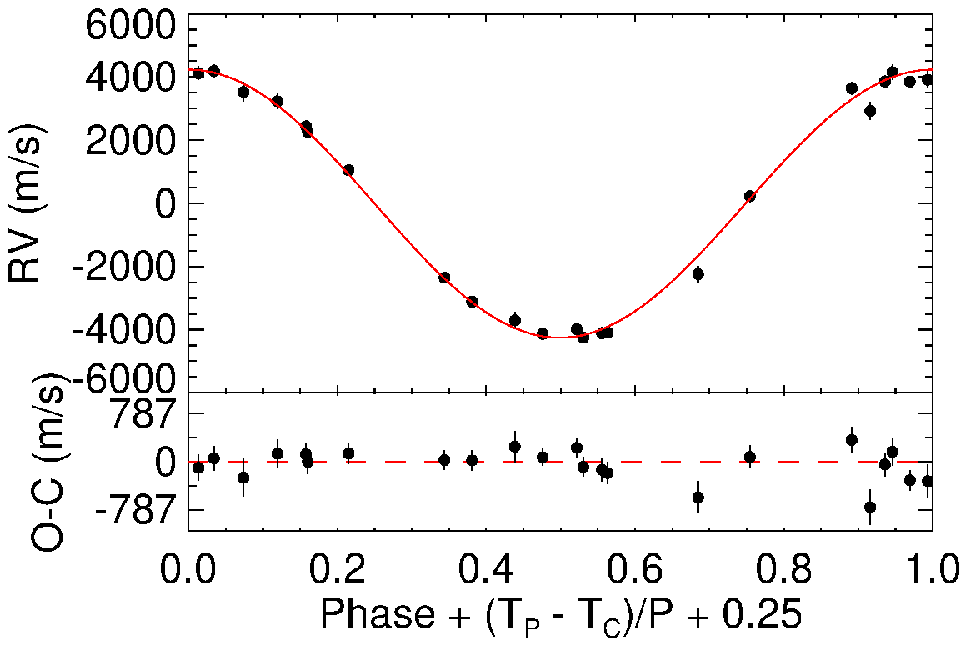}
    \caption{Archival RV observations from TrES (black data points) and the best-fitting Keplerian model (red continuous line). Residuals from the model are given in the bottom panel.}
    \label{fig:rvmodel}
\end{figure}

We then simultaneously modelled the transit light curves that we selected together with the archival radial velocity (RV) data \citep{siverd2012} and the phase-folded TESS sector-17 data, supported by the information from the stellar evolution models, to derive orbital and physical properties of the system in {\sc exofastv2} \citep{eastman2017, eastman2019}. $T_{\rm eff}$ and [Fe/H] values from high-resolution spectroscopy and $R_{\star}$ from the SED fitting were assigned to the centres of the Gaussian priors. $\log{g}$ was most precisely constrained by the mean stellar density ($\rho_{\star}$) from the transit duration, via Kepler's third law, while TESS phase curve helps to better constrain the orbital eccentricity. All parameters apart from $T_{\rm eff}$, [Fe/H] and $R_{\star}$ were assigned to uniform priors between $\pm\infty$. The initial values for these fundamental stellar parameters, on the other hand, were determined from our preliminary analysis by making use of the relations given by \citet{torres2010}, instead of using stellar evolutionary tracks, and we used only one light curve for each passband for speed in computation. The age and mass of the star ($M_{\star}$) were derived from the integrated MIST model, based on $\rho_{\star}$ and $R_{\star}$. The quadratic limb darkening coefficients were interpolated from the tables of \citet{claret2011} for each passband used in the observations, which were then set to the centres of uniform priors during the light curve modelling. Convergence is ensured for the parameters when independent chains of the MCMC run are found to be similar to each other, which is controlled by the Gelman-Rubin statistic ($Rz$), and the number of independent draws ($Tz$) being large enough so that the chains are sufficiently long compared to their correlation lengths \citep{eastman2019}. Then the posterior distribution functions for the fit parameters are representative of the underlying posterior and the fit has converged. The computations stopped when converged, and a global model of the most precise transit light curves with a wide wavelength coverage, archival RV data, and broadband apparent magnitudes of the host was achieved. The values of the fit parameters are provided in Table\,\ref{tab:globalmodel}, while the models for the light, phase and RV curves based on these parameter values are presented in Figs.\ \ref{fig:lcmodel}, \ref{fig:phasecurve} and \ref{fig:rvmodel}, respectively.

\renewcommand{\arraystretch}{1.15}
\setlength\tabcolsep{2 pt}
\begin{table}
  \centering
  \caption{Median values and 68\% confidence interval for KELT-1\,b.}
  \begin{tabular}{llc}
 \hline \hline
 Symbol & Parameter (Unit) & Values \\
 \hline
 \multicolumn{2}{l}{Stellar parameters:}&\\
 \hline
$M_{\star}$&Mass (M$_{\odot}$)&$1.369^{+0.071}_{-0.075}$\\
$R_{\star}$&Radius (R$_{\odot}$)&$1.501\pm0.036$\\
$L_{\star}$&Luminosity (R$_{\odot}$)&$3.56^{+0.22}_{-0.21}$\\
$\rho_{\star}$&Density (cgs)&$0.57^{+0.032}_{-0.028}$\\
$\log{g}$&Surface gravity (cgs)&$4.221^{+0.018}_{-0.017}$\\
$T_{\rm eff}$&Effective Temperature (K)&$6471^{+50}_{-49}$\\
$[{\rm Fe/H}]$&Metallicity (dex)&$0.095\pm0.080$\\
$[{\rm Fe/H}]_{0}$&Initial Metallicity$^{1}$ &$0.211^{+0.066}_{-0.068}$\\
$Age$&Age (Gyr)&$1.65^{+0.89}_{-0.65}$\\
\hline
\multicolumn{2}{l}{Planetary parameters:}&\\
\hline
$P_{\rm orb}$&Period (days)&{\scriptsize $1.217493996(81)$}\\
$R_{\rm p}$&Radius (R$_{\rm jup}$)&$1.138\pm0.030$\\
$M_{\rm p}$&Mass (M$_{\rm jup}$)&$27.7^{+1.0}_{-1.1}$\\
$a$&Semi-major axis (au)&$0.02494^{+0.00042}_{-0.00046}$\\
$i$&Inclination (degrees)&$85.27^{+1.10}_{-0.85}$\\
$e$&Eccentricity &$0.0055^{+0.0084}_{-0.0039}$\\
$\omega_{\star}$&Argument of periastron (degrees)&$-100^{+120}_{-130}$\\
$T_{\rm eq}$&Equilibrium temperature$^{6}$ (K)&$2421\pm28$\\
$\tau_{\rm circ}$&Tidal circularization timescale (Gyr)&$0.133^{+0.015}_{-0.013}$\\
$K$&RV semi-amplitude (m/s)&$4207\pm57$\\
$R{\rm _P}$/R$_{\star}$&Radius of planet in stellar radii &$0.07792^{+0.00031}_{-0.00034}$\\
$a/R_{\star}$&Semi-major axis in stellar radii &$3.570^{+0.065}_{-0.061}$\\
$\delta$&Fractional transit depth&$0.006072^{+0.000049}_{-0.000052}$\\
$\tau$&Ingress/egress transit duration (days)&$0.00925^{+0.00035}_{-0.00036}$\\
$T_{14}$&Total transit duration (days)&$0.11484\pm0.00035$\\
$T_{\rm FWHM}$&FWHM transit duration (days)&$0.10560\pm0.00019$\\
$b$&Transit impact parameter &$0.295^{+0.048}_{-0.064}$\\
$\tau_S$&Ingress/egress eclipse duration (days)&$0.00920^{+0.00038}_{-0.00039}$\\
$T_{14_S}$&Total occultation duration (days)&$0.1146^{+0.0013}_{-0.0019}$\\
$A_{\rm ellips}$&Ellipsoidal variation amplitude (ppm)&$398\pm19$\\
$\delta_{S}$&Measured eclipse depth (ppm)&$388\pm53$\\
$\rho_{\rm p}$&Density (cgs)&$23.3^{+1.6}_{-1.4}$\\
log$~g_{\rm p}$&Surface gravity &$4.724^{+0.019}_{-0.018}$\\
$\Theta$&Safronov number &$0.886^{+0.027}_{-0.025}$\\
\hline
\multicolumn{2}{l}{Wavelength-dependent parameters:}&\\
\hline
$u_{1,I}$&linear limb-darkening coeff in $I$ &$0.271\pm0.016$\\
$u_{2,I}$&quadratic limb-darkening coeff in $I$ &$0.352\pm0.018$\\
$u_{1,R}$&linear limb-darkening coeff in $R$ &$0.286\pm0.025$\\
$u_{2,R}$&quadratic limb-darkening coeff in $R$ &$0.337\pm0.031$\\
$u_{1,i'}$&linear limb-darkening coeff in $i'$ &$0.283\pm0.033$\\
$u_{2,i'}$&quadratic limb-darkening coeff in $i'$ &$0.346\pm0.042$\\
$u_{1,r'}$&linear limb-darkening coeff in $r'$ &$0.325\pm0.035$\\
$u_{2,r'}$&quadratic limb-darkening coeff in $r'$ &$0.342\pm0.044$\\
$u_{1,z'}$&linear limb-darkening coeff in $z'$ &$0.216\pm0.043$\\
$u_{2,z'}$&quadratic limb-darkening coeff in $z'$ &$0.336\pm0.047$\\
$u_{\rm 1,TESS}$&linear limb-darkening coeff in TESS &$0.179\pm0.032$\\
$u_{\rm 2,TESS}$&quadratic limb-darkening coeff in TESS &$0.257\pm0.042$\\
$u_{1,V}$&linear limb-darkening coeff in $V$ &$0.349\pm0.037$\\
$u_{2,V}$&quadratic limb-darkening coeff in $V$ &$0.312\pm0.044$\\
\hline
\label{tab:globalmodel}
\end{tabular}
\end{table}

\subsection{Transit timing analysis}
\label{sec:ttv_analysis}

In order to investigate a potential nonlinearity in the transit timings of KELT-1\,b, we collected all the available transit light curves for the target in the Exoplanet Transit Database (ETD)\footnote{http://var2.astro.cz/ETD/}, literature, and the TESS DVT light curves. We converted the mid-exposure timings of all data points to BJD$_{\rm TDB}$ if they were provided in a different time convention, based on the location of the telescope, the coordinates of the object and the reported timing, using the relevant modules in {\sc astropy} \citep{astropy2013, astropy2018}. We had to correct some of the light curves for linear trends primarily due to the effect of airmass. 

In order to calculate mid-transit times from all of the available light curves, we used an approach that aims at homogeneity in the analysis. We derived the times of mid-transit from the {\sc exofast} transit models of all the light curves. We made use of the web-based version of the code through NASA Exoplanet Archive, which relies on the {\sc amoeba} optimization algorithm. The transit parameters set as free and their initial values gathered from \cite{siverd2012} This modelling choice significantly reduced the time spent in computations of the mid-transit times.

Although a probabilistic modelling scheme to derive the distributions for system parameters and their uncertainties would be optimal ({\sc exofastv2}), the accuracy and precision of the mid-transit timings from a fitting procedure based on an optimization algorithm ({\sc exofastv1}) are still reliable. We compared the mid-transit timings we measured using the web-based version of {\sc exofastv1} to those from our global modelling with {\sc exofastv2}, and found that the difference is less than 20\,s and within $1.3~\sigma$ of all measurements having a minimum error bar of 15\,s. This resulted in a homogeneously measured set of 72 mid-transit timings for KELT-1\,b within a significantly reduced computation time.

In order to compute the timings expected from a linear ephemeris, we employed the orbital period ($P_{\rm orb}$) value from \cite{maciejewski2018} and the mid-transit time ($T_0$) we measured from the most precise light curve that we acquired with the 1.23\,m telescope at CAHA on 20 August 2015 as reference elements. We then calculated the differences between these expected mid-transit timings based on a linear ephemeris (denoted as C for calculated from ${\rm C} = T_0 + {\rm E} \times P_{\rm orb}$) and the actual observations of the timings of the conjunctions (denoted as O for observed) and plotted them with respect to the orbital cycle (epoch, E), and hence formed an O-C diagram. We had to eliminate only two data points, both from the ETD, because they deviated in the diagram by more than 3$\sigma$ of the mean O-C. These issues were very unlikely to be caused by an astrophysical mechanism that would lead to a sudden shift of these transits. This resulted in a total of 72 precise mid-transit times for analysis. A linear trend with a small slope was found due to the accumulation of uncertainties on the reference elements with time. We fitted a linear and a quadratic model to this dataset independently to update the linear ephemeris and search for a secular change in the orbital period, respectively.

We generated random samples for the fit parameters using a Markov Chain Monte Carlo (MCMC) algorithm based on 16 chains, 5000 iterations and a burn-in period to discard the first 500 steps, by making use of the relevant functions in the PyMC3 package \citep{pymc3} and calculated the likelihood of each sample based on its agreement with the TTV diagram. The posterior probability distribution of each fit parameter was computed, its median value was taken to be the value for the corresponding parameter, and the 16th and 84th percentiles were taken to be its uncertainty. The two parameters of the linear model were used to update the linear ephemeris as provided in Eq.~\ref{eq:ephemeris}.

\begin{equation}
  T_0 =  2457254.624084(54) + 1.21749385(4) \times E
  \label{eq:ephemeris}
\end{equation}

We followed the same procedure in fitting a quadratic function to investigate secular changes in the transit timings. We calculated the likelihoods of three coefficients of a parabola sampled from an MCMC run with the same hyper-parameters as we employed during the linear fit. We found a negative quadratic coefficient, in agreement with the expected form of orbital decay. The quadratic ephemeris we found is:

\begin{equation}
 \label{eq:quadratic}
\begin{split}
  T_0 = 2457254.624150(62) + 1.21749398(7) \times E \\+ (-9.9\pm4.5)~10^{-11} \times E^2 
\end{split}
\end{equation}

\begin{figure*}
	\includegraphics[width=0.9\paperwidth]{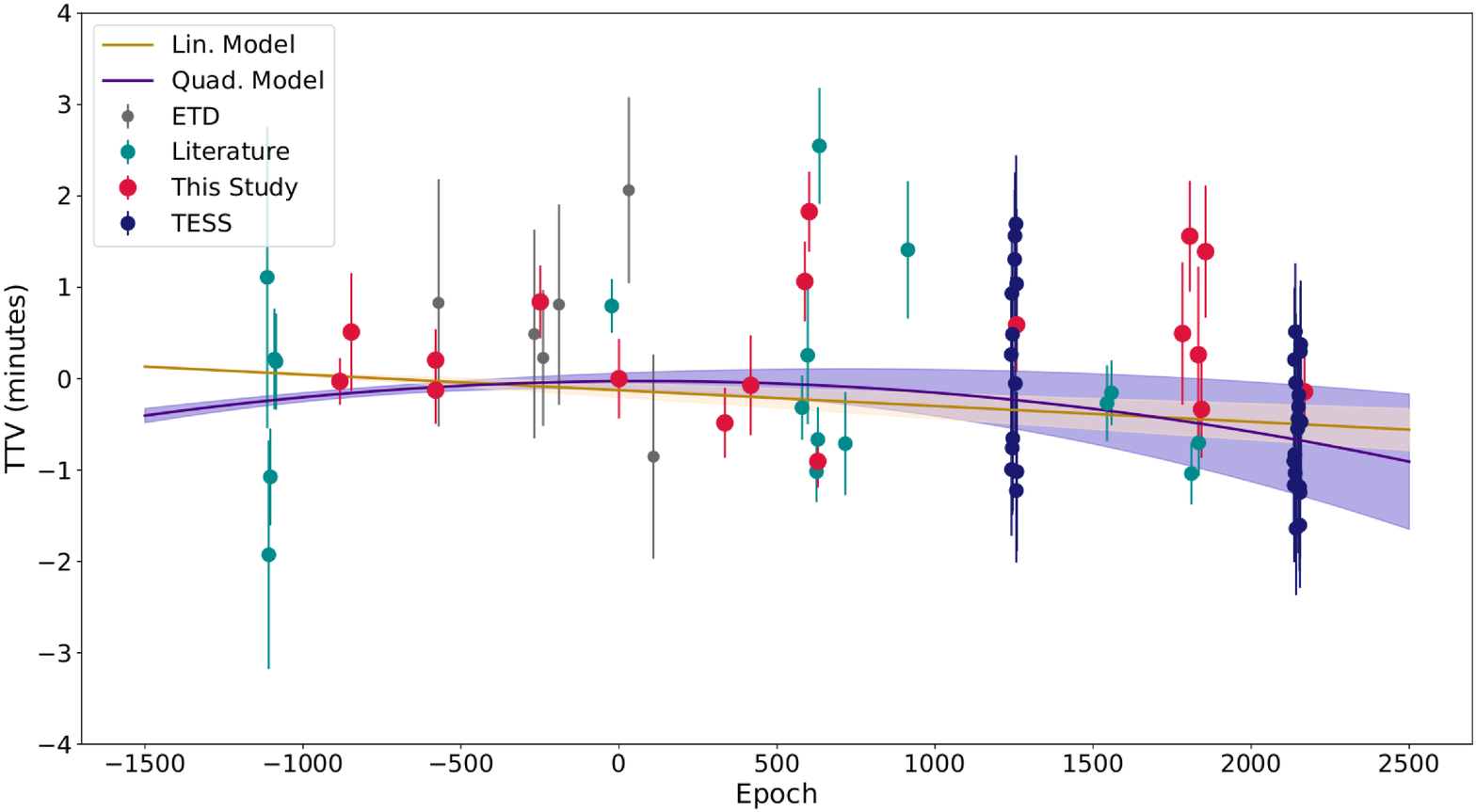}
    \caption{TTV diagram of KELT-1\,b with linear (in orange) and quadratic models (purple) superimposed on data together with their uncertainties in shades with the same colours as the models.}
    \label{fig:kelt1_ttv_diagram}
\end{figure*}

When we compare the fits, although the quadratic model performs slightly better than the linear in representing the data with a $\chi^2_{\nu}$ value of 2.273 compared to 2.308 for the latter, the values for Bayesian and Akaike Information Criteria (BIC and AIC) favour the linear model  with $\Delta$BIC and $\Delta$AIC values of 4.24 and 1.96, respectively. Since the two-parameter linear model is simpler as well, there is no evidence for an orbital decay at the moment. Nevertheless, we provide both models and their uncertainties superimposed on O-C data in Fig.~\ref{fig:kelt1_ttv_diagram}. Based on our updated ephemeris, we did not find any difference between the average mid-transit timings of two TESS sectors within the propagated uncertainties. This implies that there is no change on the orbital period over 2.9 year time interval covered by the TESS data within the limits of measurement uncertainties.

Since the linear model better fits the data, we are only able to place a lower limit on the reduced tidal quality factor. To do this we used the fifth percentile of the quadratic coefficient, $A$, to give an upper limit on the rate of change of period with observing epoch according to
\begin{equation}
A = \frac{1}{2}\frac{dP}{dE}
\end{equation}
and converted this to a limit on $Q^{\prime}_{\star}$ using the equation
\begin{equation}
    Q^{\prime}_{\star} = -\frac{27}{2} \pi \frac{M_p}{M_{\star}} \left(\frac{a}{R_{\star}}\right)^{-5} \left(\frac{dP}{dE}\right)^{-1} P_{\rm orb}
  \label{eq:tidal_quality_parameter}
\end{equation}
where the other parameters were taken from our global modelling results. Hence, assuming the system is not synchronized, we find $Q^{\prime}_{\star} > (8.5 \pm 3.9) \times 10^6$ slightly larger than the canonical value of $10^6$.

We then continued our analysis with the residuals of the O-C values from the linear model to search for potential periodic signals in the data, which would hint at the existence of gravitationally bound perturbers. We obtained a Lomb-Scargle periodogram of the residuals with the help of a Python code based on the {\sc timeseries} function of the {\sc astropy} package \citep{vanderPlas2018}. We did not find any statistically significant peak down to a $20\%$ false alarm probability (FAP) level based on mid-transit timing data from 72 light curves unevenly distributed over $\sim10.9$ years with a mean error of $41 \pm 16$\,s. Therefore we find no evidence for a significant periodicity in the TTV diagram. We followed the same analysis approach by making use of the published measurements of mid-transit timings instead of our own values, but the scatter on the O-C diagram increased and the goodness-of-fit statistics decreased significantly. This occurred despite more than two thirds (50 of 72) of the data points being our own measurements from our observations and the TESS light curves. Therefore, we continued our analysis based on the transit times we calculated in a homogeneous way. 

We experimented by making use of 49 points within $1\sigma$ of the average error bar of mid-transit timing measurements. This eliminated all the ETD data, ten light curves from the literature (including the first two light curves from \citealt{siverd2012}), and eight of our own light curves. The linear model was again found to be better than the quadratic model, and there was no significant peak in the frequency analysis. Therefore, the selection of data based on the size of the error bars did not bring any improvement except for the appearance of a candidate frequency at $\sim283$ days with $1.9\%$ FAP, which we note for future reference.

We then used the $\beta$-factor as an elimination criterion to work with the light curves least affected by correlated noise ($\beta < 2.5$). While all ETD and literature light curves passed this criterion, ten TESS transits were eliminated of which seven were from sector 17. The total scatter of the mid-transit timings from this sector is slightly larger (2.95\,min compared to 2.12\,min for sector 57) as observed from the TTV diagram in Fig.~\ref{fig:kelt1_ttv_diagram}. Therefore, we find the scatter in the TTV diagram, in excess of that implied by observational uncertainties, to be attributable to potential activity-induced spot-crossing events during transits. Spot-induced asymmetries in some of the light curves are just about noticeable. However, they are not persistent from one transit light curve to another so we cannot use them to track and recover a rotation period. The ground-based light curves, on the other hand, are heavily influenced by white noise, which dominates the red noise. When we analyze only the light curves with PNR values less than 2.0 (61 in total), all 32 TESS light curves survive. However, this choice brings no improvement to the results either. After all these experiments, we decided to present our results based on the entire data set, which we provided as online material.

\section{Conclusions}
\label{sec:conclusion}
\begin{figure}
	\includegraphics[width=0.975\columnwidth]{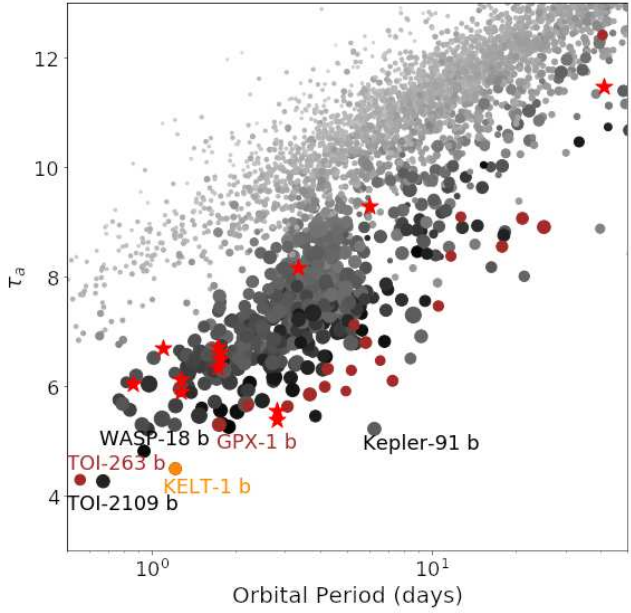}
    \caption{Plot of the $\tau_a$ parameter, which is proportional to the e-folding timescale of orbital decay, with respect to the orbital period for planets, close-in brown dwarfs and some low-mass stars in binary systems. The sizes of the data points depend on the radii of the objects (larger points for larger objects) and their colours depend on their masses (darker grey for more massive objects). BD-mass objects are in brown, also scaled with their radii, and low-mass stars are given with red stars while KELT-1 is in orange based on the parameters derived within this study, the error bars of which are smaller than the size of the marker.}
    \label{fig:demographics_dimqnt1}
\end{figure}

KELT-1\,b was one of only seven transiting BDs (13 M$_{\rm Jup} < M_{\rm p} < 80$ $M_{\rm Jup}$) known at the time it was discovered \citep{siverd2012}. In the intervening ten years, this number has increased to 24 (as reported in the TEPCat\footnote{https://www.astro.keele.ac.uk/jkt/tepcat/} catalogue; \citealt{southworth2011}). This is due to the mechanisms of their formation and migration and is despite the biases brought by the transit method, which favours the detection of large objects at short orbital periods. The recently-discovered objects GPX-1\,b \citep{benni2021}, TOI-263\,b \citep{palle2021}, and TOI-519\,b \citep{parviainen2021} are other examples of BD-mass objects on short-period orbits. Considering the paucity of such massive objects in the close vicinity of their host stars (the ``brown dwarf desert''), it is reasonable to expect their orbits to decay under strong tidal interactions with their host stars. In fact, KELT-1\,b is one of the strongest candidates for orbital decay based on its e-folding timescale, which is proportional to the quantity 
\begin{equation}
\tau_a = \frac{M_{\star}}{M_{\rm p}} \left(\frac{a}{R_{\star}}\right)^5
\end{equation}
as defined by \citet{siverd2012}. We plot this dimensionless quantity versus orbital period in Fig.~\ref{fig:demographics_dimqnt1} and label some interesting objects together with KELT-1\,b. Planet data (grey circles) are based on relevant parameter values for NASA Exoplanet Archive\footnote{https://exoplanetarchive.ipac.caltech.edu/docs/data.html} composite data set, while that for BDs (brown circles) are from TEPCat. We also included 14 low-mass stellar companions with orbital periods smaller than 100\,d and less massive than 0.25 M$_{\odot}$ (red stars) from the Detached Eclipsing Binary Catalogue (DEBCat)\footnote{https://www.astro.keele.ac.uk/jkt/debcat/} \citep{southworth2015} . KELT-1\,b's location is based on the parameters derived within this study. The so-called sub-Jovian desert is visible as the relatively empty triangular area roughly between $\tau_a \sim10^7$ and $\tau_a \sim10^8$ below an orbital period of 2.5\,d \citep{basturk2020}.

Although KELT-1\,b stands out as one of the most promising candidates for period decay, planetary mass and orbital period are not the only factors determining the stability of planet's orbit. Despite being the only case for which orbital decay has been detected convincingly, WASP-12\,b does not stand out on this plot. Nevertheless, it has a relatively lower $\tau_a$ value of $\sim10^{5.4}$. 

\begin{figure}
	\includegraphics[width=0.975\columnwidth]{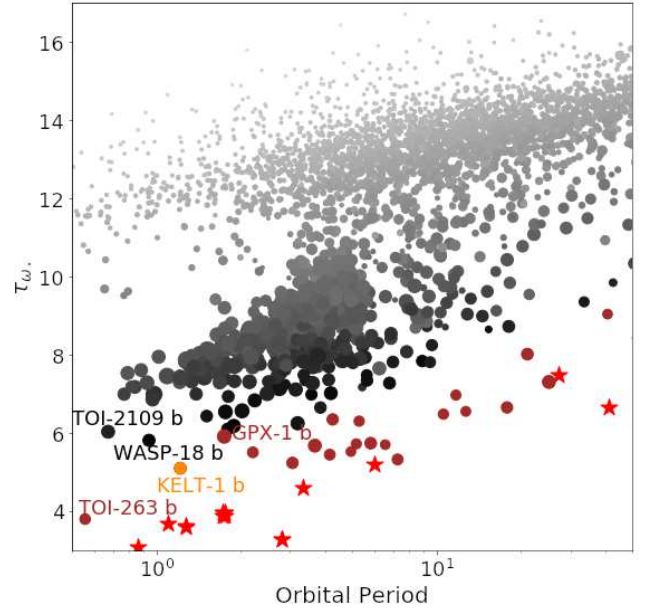}
    \caption{Plot of the $\tau_{\omega_{\star}}$ parameter, which is proportional to the synchronization timescale of the stellar spin with the orbital period, versus the orbital period of the companions. Point sizes and colour scale are the same as in Fig.~\ref{fig:demographics_dimqnt1}.}
    \label{fig:demographics_dimqnt2}
\end{figure}

Another dimensionless quantity defined by \citet{siverd2012} as 
\begin{equation}
\tau_{\omega_{\star}} = \left(\frac{M_{\star}}{M_{\rm p}}\right)^2 \left(\frac{a}{R_{\star}}\right)^3
\end{equation} 
is proportional to the synchronization timescale of the stellar spin with the orbital period. KELT-1\,b should have the shortest synchronization timescale based on its location on this plot (Fig.~\ref{fig:demographics_dimqnt2}). Tidal interactions with the planet spin up the host star and a stable tidal equilibrium can be achieved when the total angular momentum is sufficient \citep{ogilvie2014}. However, for the mid-F star hosting KELT-1\,b, the dissipation mechanism is unclear because it should have a shallow convective zone. 

Based on our analysis of timing data spanning almost eleven years of observations, we find no evidence for decay in the orbit of KELT-1\,b because the linear-ephemeris model turned out to be superior to the quadratic model. Thus, we constrained a lower limit for the tidal quality parameter as $Q^{\prime}_{\star} > (8.5 \pm 3.9) \times 10^6$. Considering that the host star is relatively hot ($T_{\rm eff} \sim6471$ K) and therefore should have a thin convective envelope, the tidal energy it should dissipate in one revolution of the planet in its tidal bulge should even be lower than that predicted by the canonical value for stars with convective envelopes. The characteristic timescale for orbital decay ($\tau_{\rm decay}$) is not more than 140\,Myr (95\% confidence) as well. However, we find an age of $1.65^{+0.89}_{-0.65}$ Gyr from isochrone fitting. We interpret these results as indicators of a tidal equilibrium in the system. Our interpretation is further supported by the proximity of the rotation periods found both from the projected rotational velocity from spectral line broadening (1.33\,d, \citealt{siverd2012}) and spot-induced modulation from the frequency spectrum (1.52\,d, \citealt{vonEssen2021}) to the orbital period (1.21\,d, this study) despite the large uncertainty on the rotational period.

Although tidal equilibrium is likely, the mechanism causing the angular momentum transfer so that the total angular momentum is sufficient for the equilibrium is unclear. Nevertheless, other similarly short-period and massive objects have been found orbiting relatively hotter stars than the Sun with thinner convective envelopes as KELT-1. In fact, more massive companions are suggested to be hosted by hotter stars on average \citep{jiang2021}. This is most probably the consequence of a more massive protoplanetary disk that would form around a larger star to support the formation of more massive planets in turn \citep{andrews2013}. This mass-mass correlation agrees with the predictions of the in situ formation mechanism \citep{armitage2018}. Then such massive planets and BDs may have formed closer to their stars by accreting material from a massive protoplanetary disk and then migrated inwards even further by interacting with it or through Kozai-Lidov oscillations and then stabilized on short-period orbits after a tidal equilibrium is achieved. After all, we still do not know either the maximum mass of the object that can be formed by accretion in a disk or the minimum mass formed by disk instability. Therefore, a classification based on its formation mechanism as a planet via core-accretion or as a BD via disk-instability is a matter of debate. Our results show that KELT-1\,b is similar to other massive planets in terms of its orbital properties. However, our global model of the available light curves (including that of TESS) and RV data with a full-phase coverage suggests an orbital eccentricity, with a significance of $\sim1.5 \sigma$, as has been noticed in earlier work \citep{beatty2014,vonEssen2021}. Since the circularization timescale is very short, a potential third body on a larger orbit can explain the small but a non-zero eccentricity if it is real, which requires more and precise follow-up transit and occultation observations of the target. They will also help us further constrain the tidal quality parameter of its host star and understand the extent of tidal interactions in this important system.

\section*{Acknowledgements}

We gratefully acknowledge the support by The Scientific and Technological Research Council of Türkiye (T\"UB\.{I}TAK) with the project 118F042. We thank T\"UB\.{I}TAK for the partial support in using T100 telescope with the project number 19AT100-1471. This work is also supported by the research fund of Ankara University (BAP) through the project 18A0759001. This research has made use of data obtained using the ATA50 telescope and CCD attached to it, operated by Atat\"urk University Astrophysics Research and Application Center (ATASAM). Funding for the ATA50 telescope and the attached CCD have been provided by Atat\"urk University (P.No. BAP-2010/40) and Erciyes University (P.No. FBA-11-3283) through Scientific Research Projects Coordination Units (BAP), respectively. This paper is also based on the observations performed with the Zeiss 1.23\,m telescope at the Centro Astron\'{o}mico Hispano Alem\'{a}n (CAHA) in Calar Alto (Spain) and the Cassini 1.52\,m telescope at the Astrophysics and Space Science Observatory of Bologna in Loiano (Italy). We thank Roberto Gualandi for his technical assistance at the Cassini telescope. We thank the support astronomers of CAHA for their technical assistance at the Zeiss telescope.
L.\,M. acknowledges support from the ``Fondi di Ricerca Scientifica d'Ateneo 2021'' of the University of Rome ``Tor Vergata''. Some/all of the data presented in this paper were obtained from the Multimission Archive at the Space Telescope Science Institute (MAST). STScI is operated by the Association of Universities for Research in Astronomy, Inc., under NASA contract NAS5-26555. Support for MAST for non-HST data is provided by the NASA Office of Space Science via grant NAG5-7584 and by other grants and contracts. This research has made use of the NASA Exoplanet Archive, which is operated by the California Institute of Technology, under contract with the National Aeronautics and Space Administration under the Exoplanet Exploration Program. This work presents results from the European Space Agency (ESA) space mission Gaia. Gaia data are being processed by the Gaia Data Processing and Analysis Consortium (DPAC). Funding for the DPAC is provided by national institutions, in particular the institutions participating in the Gaia MultiLateral Agreement (MLA). We thank all the observers who report their observations to Exoplanet Transit Database (ETD).

\section*{Data Availability}
Some of the light curves to derive mid-transit times were downloaded from Exoplanet Transit Database at http://var2.astro.cz/ETD/. All other light curves appearing for the first time in this article are presented as online material. Mid-transit times derived from our own light curves as well as that of other observers' and TESS light curves are presented as online data sets too through Vizier Online.



\bibliographystyle{mnras}
\bibliography{kelt1_rev2_ozbasturk_mnras} 








\bsp	
\label{lastpage}
\end{document}